\documentstyle[aps,epsfig]{revtex}
\begin{document}
\draft
%\preprint{Nordita }
\baselineskip=18pt

%\begin{title}
\title
 {The Electrical Conductivity in the Early Universe}
%\end{title}
\author{Gordon Baym}
%\begin{instit}
\address
      {University of Illinois at Urbana-Champaign,\\
      1110 W. Green St., Urbana, IL 61801, USA}
%\end{instit}
\author{Henning Heiselberg}
%and
%\begin{instit}
\address{      NORDITA,
      Blegdamsvej 17, DK-2100 Copenhagen \O, Denmark}
%\end{instit}
%\today
\maketitle

\begin{abstract}

    We calculate the electrical conductivity in the early universe at
temperatures below as well as above the electroweak vacuum scale, $T_c\simeq
100$GeV.  Debye and dynamical screening of electric and magnetic interactions
leads to a finite conductivity, $\sigma_{el}\sim T/\alpha\ln(1/\alpha)$, at
temperatures well below $T_c$.  At temperatures above, $W^\pm$ charge-exchange
processes -- analogous to color exchange through gluons in QCD -- effectively
stop left-handed charged leptons.  However, right-handed leptons can carry
current, resulting in $\sigma_{el}/T$ being only a factor $\sim
\cos^4\theta_W$ smaller than at temperatures below $T_c$.

\end{abstract}
%\pacs{PACS numbers: 12.38.Mh, 12.38.Bx, 52.25.Dg }
%\setlength{\oddsidemargin}{1.5cm}
%\setlength{\evensidemargin}{\oddsidemargin}
%\columnsep -.8cm
%\twocolumn
%\narrowtext
%\pagestyle{empty}

\section{Introduction}

    The strong magnetic fields measured in many spiral galaxies, $B\sim
2\times 10^{-6}$ G \cite{B}, are conjectured to be produced primordially;
proposed mechanisms include fluctuations during an inflationary universe
\cite{Turner} or at the GUT scale \cite{EO}, and plasma turbulence during the
electroweak transition \cite{Hogan,LMcL,olinto1} or in the quark-gluon
hadronization transition \cite{qcd,olinto}.  The production and later
diffusion of magnetic fields depends crucially on the electrical conductivity,
$\sigma_{el}$, of the matter in the universe; typically, over the age of the
universe, $t$, fields on length scales smaller than $L\sim
(t/4\pi\sigma_{el})^{1/2}$ are damped.  In this paper we calculate
$\sigma_{el}$ in detail below and above the electroweak transition scale.

    The electrical conductivity was estimated in \cite{Turner} in the
relaxation time approximation as $\sigma_{el}\sim n\alpha\tau_{el}/m$ with
$m\sim T$ and relaxation time $\tau_{el}\sim 1/(\alpha^2T)$, where $\alpha =
e^2/4\pi$.  In Refs.  \cite {HK} and \cite{Enqvist} the relaxation time was
corrected with the Coulomb logarithm.  A deeper understanding of the screening
properties in QED and QCD plasmas in recent years has made it possible to
calculate a number of transport coefficients including viscosities, diffusion
coefficients, momentum stopping times, etc., exactly in the weak coupling
limit \cite{BP,tran,eta}; also \cite{GT}.  However, calculation of processes
that are sensitive to very singular forward scatterings remain problematic.
For example, the calculated color diffusion and conductivity \cite{color},
even with dynamical screening included, remain infrared divergent due to color
exchange in forward scatterings.  Also the quark and gluon damping rates at
non-zero momenta calculated by resumming ring diagrams exhibit infrared
divergences \cite{damp} whose resolution requires more careful analysis
including higher order diagrams as, e.g., in the Bloch-Nordseick calculation
of Ref.  \cite{Blaizot}.  Charge exchanges through $W^\pm$ exchange, processes
similar to gluon color exchange in QCD, are important in forward scatterings
at temperatures above the $W$ mass, $M_W$.  While, as we show, such processes
lead negligible transport by left-handed charged leptons, they do not prevent
transport by right-handed charged leptons.  As a consequence, electrical
conduction at temperatures above the electroweak transition is large, and does
not inhibit generation of magnetic fields \cite{Enqvist}; the observed
magnetic fields in galaxies could then be generated earlier than the
electroweak transition\cite{Hogan,LMcL} and have survived until today.  More
generally, we find that the electrical conductivity is sufficiently large that
it does not lead to destruction of large-scale magnetic flux over timescales
of the expansion of the universe.

    In Sec.  II we calculate the electrical conductivity for $T\ll M_W$.  In
this regime the dominant momentum transfer processes in electrical transport
are electrodynamic, and one can ignore weak interactions.  We turn then in
Sec.  III to the very early universe, $T\gg T_c$, where the $W^\pm$ are
effectively massless and their effects on electrical conduction must be taken
into account.

\section{Electrical conductivities in high temperature QED}

    We first calculate the electrical conductivity in the electroweak
symmetry-broken phase at temperatures well below the electroweak boson mass
scale, $T\ll M_W$.  As we argue below, charged leptons $\ell=e^-,\mu^-,\tau^-$
and anti-leptons $\bar{\ell}=e^+,\mu^+,\tau^+$ dominate the currents in the
regime in which $m_{\ell}\ll T$.  In the broken-symmetry phase weak
interactions between charged particles, which are generally smaller by a
factor $\sim(T/M_W)^4$ compared with photon-exchange processes, can be
ignored.  The primary effect of strong interactions is to limit the drift of
strongly interacting particles, and we need consider only electromagnetic
interactions between charged leptons and quarks.

    Transport processes are most simply described by the Boltzmann kinetic
equation for the distribution functions, $n_i({\bf p},{\bf r},t)$, of particle
species $i$, of charge $e$,
\begin{eqnarray}
   \left(\frac{\partial}{\partial t} + {\bf v}_1\cdot
   \nabla_{\bf r}+ e{\bf E}\cdot\nabla_1 \right) n_1
   = -2\pi \nu_2\sum_{234}&|M_{12\to 34}|^2
   [n_1n_2(1\pm n_3)(1\pm n_4)-n_3n_4(1\pm n_1)(1\pm n_2)]
   \nonumber \\
   &\times\delta_{{\bf p}_1+{\bf p}_2, {\bf p}_3+{\bf p}_4}
   \delta(\epsilon_1 +\epsilon_2 -\epsilon_3 -\epsilon_4 ),
   \label{BE}
\end{eqnarray}
where ${\bf E}$ is the external electric field driving the charges, and
the right side describes the two-particle collisions $12\leftrightarrow 34$
slowing them down.  The $\pm$ signs refer to bosons and fermions.  The sums
are over momenta ${\bf p}_2$, ${\bf p}_3$, and ${\bf p}_4$, and the
statistical factor $\nu_2$ accounts for the number of leptons and spin
projections that scatter with particle 1. Massless lepton-lepton or
antilepton-antilepton scattering conserves the electrical current, and affects
the conductivity only in higher order.  In lowest order we need to take into
account only lepton-antilepton scattering.  The square of the matrix element
for scattering of a charged lepton from initial momentum ${\bf p}_1$ to final
momentum ${\bf p}_3$ with an antilepton from ${\bf p}_2$ to ${\bf p}_4$ is
\begin{eqnarray}
|M_{12\to 34}|^2=2e^4(s^2+u^2)/t^2/
(16\epsilon_1\epsilon_2\epsilon_3\epsilon_4),
 \label{matrix}
\end{eqnarray}
where $s,t,u$ are the usual Mandelstam variables formed from the incoming
and outgoing momenta.

    To solve the kinetic equation for a weak driving field, we write $n_i =
n_i^0 + \Phi_i$, where $n^0_i=(\exp(\epsilon_i/T)\pm1)^{-1}$ is the global
equilibrium distribution (chemical potentials are taken to vanish in the early
universe), and linearize Eq.  (\ref{BE}) in the deviations $\Phi_i$ of the
distribution functions from equilibrium (see Refs.  \cite{BPbook}, \cite{BHP}
and \cite{eta} for details); thus Eq.  (\ref{BE}) reduces to
\begin{eqnarray}
  &e{\bf E}&\cdot {\bf v}_1 \frac{\partial n_1}{\partial\epsilon_1} =
 \, -2\pi \nu_2\sum_{234} |M_{12\to 34}|^2
 n^0_1n^0_2(1\pm n^0_3)(1\pm n^0_4)
\left(\Phi_1+\Phi_2-\Phi_3-\Phi_4\right)
% \nonumber \\ &\times&
\delta_{{\bf p}_1+{\bf p}_2, {\bf p}_3+{\bf p}_4}
\delta(\epsilon_1 +\epsilon_2 -\epsilon_3 -\epsilon_4 ).
\label{BE2}
\end{eqnarray}

    We set the stage by considering the simple situation of conduction by
massless charged leptons (component 1) and antileptons (component 2) limited
by their scattering together; later we include effects of quarks.  The
electric field drives the charged leptons and antileptons into flow with
steady but opposite fluid velocities ${\bf u}_1=-{\bf u}_2$ in the
center-of-mass system (see Fig. 1).  Assuming that collisions keep the
drifting components in local thermodynamic equilibrium, we approximate the
quasiparticle distribution functions by those of relativistic particles in
equilibrium with a local flow velocity ${\bf u}$,
\begin{eqnarray}
   n_i({\bf p}_i) =\frac{1}{\exp[(\epsilon_i-{\bf
     u}_i\cdot{\bf p}_i)/T]\mp 1};
     \label{n}
\end{eqnarray}
the deviation $\Phi_i$ is thus
\begin{eqnarray}
   \Phi_i = -{\bf u}_i\cdot {\bf p}_i \frac{\partial n_i}{\partial\epsilon_i}.
   \label{Phi}
\end{eqnarray}
The ansatz (\ref{n}) for the distribution function is an excellent
approximation; a more elaborate variational calculation which we have carried
out (see \cite{eta} for analogous calculations for the viscosity) gives almost
the same distribution function, and a corresponding electrical conductivity
only 1.5\% smaller than our result, Eq.  (\ref{ec}) below.  Equation
(\ref{Phi}) gives the total electric current from lepton-antilepton pairs with
$N_\ell$ effectively massless species present at temperature $T$:
\begin{eqnarray}
    {\bf j}_{\ell\bar{\ell}} = e n_{\rm ch}{\bf u}_1
    = e{\bf u}_1 \frac{3\zeta(3)}{\pi^2}N_\ell T^3,
\end{eqnarray}
where $n_{ch}$ is the number density of electrically charged leptons plus
antileptons.  Note that since photon exchange does not transfer charge (as do
$W^\pm$), particle 3 in Eq.  (\ref{BE}) has the same charge as particle 1, and
particle 4 the same as 2, and ${\bf u}_3={\bf u}_1$ and ${\bf u}_4={\bf
u}_2=-{\bf u}_1$.

    To calculate $\sigma_{el}$ we multiply Eq.  (\ref{BE2}) by $\nu_1{\bf v}_1$
and sum over ${\bf p}_1$, to find an equation of the form:
\begin{eqnarray}
-e{\bf E}N_\ell T^2/18 = -\xi {\bf u_1}
  \label{Xi}
\end{eqnarray}
where $\xi$ results from the right side of (\ref{BE2}).  Since QED
interactions are dominated by very singular forward scattering arising from
massless photon exchange, the sum on the right side in $\xi$ diverges unless
we include Debye screening of longitudinal interactions and dynamical
screening of transverse interactions due to Landau damping.  This is done by
including (in the Coulomb gauge) the Coulomb self-energy, $\Pi_L$, and the
transverse photon self-energy, $\Pi_T$, in the longitudinal and transverse
electromagnetic propagators.  The inverse of the longitudinal propagator
becomes $q^2+\Pi_L(\omega,q)$, and the inverse of the transverse propagator,
$t=\omega^2-q^2$ becomes $\omega^2-q^2-\Pi_T(\omega,q)$, where $\omega$ and
$q$ are the energy and momentum transferred by the electromagnetic field in
the scattering.  The quantity $\xi$ can be calculated to leading logarithmic
order in the coupling constant by expanding for small momentum transfers (see
Ref.  \cite{eta} for details).  Small momentum transfer processes are screened
by $\Pi_L\sim q_D^2=4\pi\alpha N_l T^2/3$, and $\Pi_T\sim i\pi q_D^2\omega/
4q$.  (Large momentum transfers, $q\raisebox{-.5ex}{$\stackrel{>}{\sim}$}
 \langle p\rangle \sim 3T$, are cut off
by the distribution functions.)  The resulting integrals, $\int
q^2d^2q/|q^2+\Pi_{L,T}|^2$, give characteristic logarithms, $\ln(T/q_D)$, and
we find,
\begin{eqnarray}
\xi = \frac{2\ln2}{9\pi}N_{\ell}^2\alpha^2\ln(C/\alpha N_l) T^4.
 \label{Xi2}
\end{eqnarray}
The constant $C\sim 1$ in the logarithm gives the next to leading order
terms (see \cite{eta} for the calculation of second-order contributions to the
viscosity).  The electrical conductivity for charged leptons is thus
\cite{tran,BHP}
\begin{eqnarray}
   \sigma_{el}^{(\ell\bar{\ell})} &\equiv& j_{\ell\bar{\ell}}/E =\,
\frac{3\zeta(3)}{\ln2} \frac{T}{\alpha\ln(1/\alpha N_l)}\,,
     \quad m_e\ll T\ll T_{QCD}.
     \label{ec}
\end{eqnarray}
Note that the number of lepton species drops out except in the logarithm.
The above calculation taking only electrons as massless leptons ($N_l=1$)
gives a first approximation to the electrical conductivity in the temperature
range $m_e\ll T\;\ll\; T_{QGP}$, below the hadronization transition,
$T_{QGP}\sim 150$ GeV, at which hadronic matter undergoes a transition to a
quark-gluon plasma.  Thermal pions and muons in fact also reduce the
conductivity by scattering electrons, but they do not become significant
current carriers because their masses are close to $T_{QGP}$.

    For temperatures $T > T_{QGP}$, the matter consists of leptons and
deconfined quarks.  The quarks themselves contribute very little to the
current, since strong interactions limit their drift velocity.
 The quark drift velocity can be estimated by replacing
$\alpha$ by the strong interaction fine structure constant, $\alpha_s$, in
$\xi$, Eqs.  (\ref{Xi}) and (\ref{Xi2}), which yields ${\bf u}_q\sim {\bf
u}_\ell (\alpha^2\ln\alpha^{-1})/(\alpha_s^2\ln\alpha_s^{-1})$.

    Even though quarks do not contribute significantly to the currents they
are effective scatterers, and thus modify the conductivity (an effect ignored
in the recent numerical analysis of Ref.  \cite{Ahonen}).  To calculate the
quark contribution to the lepton conductivity, we note that the collision term
between leptons (1,3) and quarks (2,4) includes the following additional
factors compared with lepton-lepton scattering:  a factor 1/2, because the
quark velocity, ${\bf u}_2$, is essentially zero, a factor 3 from colors, and
a factor 2 because the charged leptons collide on both $q$ and $\bar{q}$;
finally we must sum over flavors with a weight $Q_q^2$, where $Q_qe$ is the
charge of quark flavor $q=u,d,s,c,b,t$.  We must also divide by the number of
leptons, $N_l$, to take into account the number of quark scatterings relative
to lepton scatterings.  Including $\ell q$ and $\ell \bar q$ collisions on the
right side of Eq.  (\ref{ec}) we find the total electrical conductivity of the
early universe \cite{tran,BHP}:
\begin{eqnarray}
\sigma_{el} =
    \frac{N_l}{N_l+ 3 \sum_q^{N_q} Q^2_q}\sigma_{el}^{\ell\bar{\ell}}
    =\frac{N_l}{N_l+ 3 \sum_q^{N_q} Q^2_q}
      \frac{3\zeta(3)}{\ln2} \frac{T}{\alpha\ln(1/\alpha N_l)}, \quad
         T_{QGP}\ll\; T\;\ll\; M_W.
\label{eq}
\end{eqnarray}

    The charged lepton and quark numbers $N_l$ and $N_q$ count only
the species present in the plasma at a given temperature, i.e., those
with masses $m_i\raisebox{-.5ex}{$\stackrel{<}{\sim}$}T$.  Figure 2
illustrates the conductivities (\ref{ec},\ref{eq}).  For simplicity
this figure assumes that the quarks and leptons make their appearance
abruptly when $T$ becomes $>m_i$; in reality they are gradually
produced thermally as $T$ approaches their masses.  Since a range of
particle energies in the neighborhood of the temperature is included,
possible resonance effects in scatterings are smeared out.

    We will not attempt to calculate the electrical conductivity in
the range $M_W\raisebox{-.5ex}{$\stackrel{<}{\sim}$} T
\raisebox{-.5ex}{$\stackrel{<}{\sim}$} T_c$ below
the critical temperature.  Recent lattice calculations \cite{Kajantie}
predict a relatively sharp transition at a temperature $T_c\sim 100$
GeV from the symmetry broken phase at low temperatures, $T\ll T_c$, to
the symmetric phase at $T\gg T_c$.  The transition is sensitive,
however, to the unknown Higgs mass, with a first order transition
predicted only for Higgs masses below $\sim 90$GeV.  The calculations
of the conductivity are technically more involved when masses are
comparable to the temperature.  Furthermore one must begin to include
contributions of the $W^\pm$ to currents and scatterings, as the
thermal suppression of their density decreases near the transition.

\section{The symmetry-restored phase}

    To calculate the conductivity well above the electroweak transition, $T\gg
T_c$, where the electroweak symmetries are fully restored, we describe the
electroweak interactions by the standard model Weinberg-Salam Lagrangian with
minimal Higgs:
\begin{eqnarray}
   {\cal L}_{MSM} &=& -\frac{1}{4}{\bf W}_{\mu\nu}\cdot {\bf W}^{\mu\nu}
                -\frac{1}{4}B_{\mu\nu}B^{\mu\nu}
             +  \bar{L}\gamma^\mu\left(i\partial_\mu-
       \frac{g}{2}\mbox{\boldmath $\tau$} \cdot{\bf W}_\mu
       -\frac{g'}{2}YB_\mu\right)L
     + \bar{R}\gamma^\mu\left(i\partial_\mu-\frac{g'}{2}YB_\mu\right)R
          \nonumber\\
   &+& \left| \left(i\partial_\mu-
     \frac{g}{2}\mbox{\boldmath$\tau$}\cdot{\bf W}_\mu
    -\frac{g'}{2}YB_\mu\right)\phi   \right|^2
   - \mu^2\phi^\dagger\phi -\lambda(\phi^\dagger\phi)^2
   - G_1\bar{L}\phi R+iG_2\bar{L}\tau_2\phi^*R + h.c.
        \, .       \label{EW}
\end{eqnarray}
Here $L$ denotes left-handed doublet and $R$ right-handed singlet leptons
or quarks, $e=g\sin\theta_W=g'\cos\theta_W$ and electrical charge $Q=T_3+Y/2$.
The last terms provide masses for leptons and quarks in the low temperature
phase, $T<T_c$, where the Higgs field has a non-vanishing vacuum expectation
value $\langle\phi\rangle=(0,v)/\sqrt{2}$; at zero temperature
$v^2=-\mu^2/\lambda=1/(G_F\sqrt{2})=4M_W^2/g^2=(246{\rm GeV})^2$.

    At temperatures below $T_c$, where $\mu^2<0$, the Higgs mechanism
naturally selects the representation $W^\pm$, $Z^0$, and $\gamma$ of the four
intermediate vector bosons.  At temperatures above the transition -- where
$\langle\phi\rangle$ vanishes for a sharp transition, or tends to zero for a
crossover -- we consider driving the system with external vector potentials
$A^a=B,W^\pm,W^3$, which give rise to corresponding ``electric'' fields ${\bf
E}_a$, where
\begin{eqnarray}
    E_i^a &\equiv& F_{i0}^a = \partial_i A_0^a - \partial_0 A_i^a,
                                  \quad A^a=B  \label{Bdef} \\
          &\equiv& F_{i0}^a = \partial_i A_0^a - \partial_0 A_i^a
             -g \epsilon_{abc}A_i^bA_0^c, \quad A^a=W^1,W^2,W^3.
             \label{Wdef}
\end{eqnarray}
One can equivalently drive the system with the electromagnetic and weak
fields derived from $A$, $Z^0$, and $W^\pm$, as when $T\ll T_c$, or any other
rotated combination of these.  We consider here only the weak field limit and
ignore the nonlinear driving terms in Eq.  (\ref{Wdef}).  The self-couplings
between gauge bosons are important, however, in the scattering processes in
the plasma determining the conductivity, as we discuss below.\footnote{Pure
SU(2) gauge fields undergo a confining first order phase transition \cite{SU2}
at a critical temperature, $T_c$, where the fields are intrinsically strongly
interacting.  In an academic universe without Higgs fields the electroweak
phase transition is non-existent.  If, in this case, we run the temperature
from far above $M_W$ to low temperatures the electroweak running coupling
constants diverge at a very small temperature $\sim 10^{-23}$GeV, signalling
that the interactions of the fields have become strong, and that one is in the
neighborhood of the SU(2) phase transition.  For our purposes we are therefore
safe in ignoring such non-linear effects and the SU(2) phase transition.
However, the nature of electrical conduction in the confined state by charged
``mesons" formed, e.g., by $e^+\nu_e$ or $u\bar d$, remains an interesting
problem in principle.  We thank Eduardo Fradkin for calling this issue to our
attention.}

    The electroweak fields $A^b$ act on the matter to generate currents $J_a$
of the various particles present in the plasma, such as left and
right-handed leptons and their antiparticles, and quarks, vector bosons, and
Higgs bosons.  The Higgs and vector boson contributions are, as we shall see,
negligible. Therefore the significant terms in the currents are
\begin{eqnarray}
    J_B^\mu &=& \frac{g'}{2} (\bar{L}\gamma^\mu Y L
                          +\bar{R}\gamma^\mu Y R) \\
    J_{W^i}^\mu &=& \frac{g}{2} \bar{L}\gamma^\mu \tau_iL.
\label{J}
\end{eqnarray}
We define the conductivity tensor $\sigma_{ab}$ in general by
\begin{eqnarray}
    {\bf J}_a = \sigma_{ab} {\bf E}^b. \label{sdef}
\end{eqnarray}
Equation (\ref{sdef}) with the equations of motion in a gauge with
$\partial^\mu A_\mu^a = 0$ yields the weak field flux diffusion equation for
the transverse component of the fields, as in QED,
\begin{eqnarray}
 (\partial^2_t-\nabla^2){\bf A}_a = \sigma_{ab}\partial_t{\bf A}_b .
\end{eqnarray}
describing the decay of weak fields in terms of the the conductivity.

    The electroweak $U(1)\times SU(2)$ symmetry implies that the conductivity
tensor, $\sigma_{ab}$, in the high temperature phase is diagonal in the
representation $a,b=B,W^1,W^2,W^3$, as can be seen directly from the (weak
field) Kubo formula
\begin{eqnarray}
  \sigma_{ab}=- \lim_{\omega\to 0} \lim_{k\to 0} \frac{1}{\omega} {\rm Im}\,
        \langle J_aJ_b\rangle_{\rm irr},
\end{eqnarray}
which relates the conductivity to (one-boson irreducible) current-current
correlation functions.\footnote{Since the linearized currents are proportional
to the Pauli spin matrices $\tau^a$ for the $W^a$ ($a$=1,2,3) fields and the
identity matrix $\tau_0=1$ for the $B$ field, one finds explicitly in one-loop
order that $\sigma_{ab}\propto Tr\{\tau_a\tau_b\}=2\delta_{ab}$ ($a$=0,1,2,3).
Including the dominant interactions in the current-current loop is basically
equivalent to solving the Boltzmann equation, which produces no off-diagonal
elements.} The construction of the conductivity in terms of the Kubo formula
assures that the conductivity and hence the related entropy production
in electrical conduction are positive.  Then
\begin{eqnarray}
  \sigma = \left( \begin{array}{cccc}
            \sigma_{BB} & 0 & 0 & 0 \\
            0 & \sigma_{WW} & 0 & 0 \\
            0 & 0 & \sigma_{WW} & 0 \\
            0 & 0 & 0 & \sigma_{WW}
                  \end{array} \right) \,.
\end{eqnarray}
Due to isospin symmetry of the $W$-interactions the conductivities
$\sigma_{W^iW^i}$ are the same, $\equiv\sigma_{WW}$, but differ from the
$B$-field conductivity, $\sigma_{BB}$.

    The calculation of the conductivities $\sigma_{BB}$ and $\sigma_{WW}$ in
the weak field limit parallels that done for $T\ll T_c$.  The main difference
is that weak interactions are no longer suppressed by a factor $(T/M_W)^4$ and
the exchange of electroweak vector bosons must be included.  The conductivity,
$\sigma_{BB}$, for the abelian gauge field $B$ can be calculated similarly to
the electrical conductivity at $T\ll T_c$.  Taking into account the fact that
both left-handed neutrinos and charged leptons couple to the $B$-field with
the same sign, and that they scatter the same way, their flow velocities are
equal.  Consequently, in the scatterings $12\leftrightarrow 34$, ${\bf u}_1
={\bf u}_3$ and ${\bf u}_2={\bf u}_4$, whether or not the interaction is by
charge exchange.  The situation is thus similar to electrodynamic case.

    Although the quarks and $W^\pm$ are charged, their drifts in the presence
of an electric field do not significantly contribute to the electrical
conductivity.  Charge flow of the quarks is stopped by strong interactions,
while similarly flows of the $W^\pm$ are effectively stopped by $W^+ + W^- \to
Z^0$, via the triple boson coupling.  Charged Higgs bosons are likewise
stopped via $W^\pm\phi^\dagger\phi$ couplings.  These particles do, however,
affect the conductivity by scattering leptons.

    The lepton and quark mass terms in the Weinberg-Salam Lagrangian provide
masses only when the Higgs field has a non-zero expectation value.  For $T\gg
T_c$ the quarks and leptons have thermal masses, which, for the longitudinal
(electric) degrees of freedom, are of order the plasma frequency, $m_{pl}\sim
gT$, and of likely order $m_{mag}\sim g^2T$ \cite{Linde} for the transverse
(magnetic) mass.  These small masses give rise to spin-flip interactions
changing the helicity; such interactions are, however, suppressed by factors
of $m/T$, and can therefore be neglected here.  The mass terms also provide a
small coupling, $G_l=\sqrt{2}m_l/v$, between the Higgs and leptons,
proportional to the ratio of the lepton mass to the vacuum expectation value,
which leads to a negligibly small contribution to the conductivity.  Even the
coupling of $\tau$ mesons with the Higgs is a factor $G_\tau^2/e^2\sim
10^{-3}$ smaller than their coupling to the $B$ field.  Charge transfer
scatterings via Higgs exchange are more singular than scatterings via $B$
exchange, and are enhanced by a factor $1/e^2$; nonetheless such processes are
negligible.

    These considerations imply that the $B$ current consists primarily of
right-handed $e^\pm$, $\mu^\pm$ and $\tau^\pm$, interacting only through
exchange of uncharged vector bosons $B$, or equivalently $\gamma$ and $Z^0$.
Because the left-handed leptons interact through ${\bf W}$ as well as through
$B$, they give only a minor contribution to the current.  They are, however,
effective scatterers of right-handed leptons.  The resulting conductivity is
\begin{eqnarray}
   \sigma_{BB} &=& \frac{1}{2}\frac{N_l\cos^2\theta_W}
      {[\frac{1}{8}(Y_R^2+2Y_L^2)N_l
       +  \frac{1}{8}\sum_q^{N_q} (Y^2_{q_R}+Y^2_{q_L})]}
      \sigma_{el}^{(\ell\bar{\ell})}
     = \frac{9}{19} \cos^2\theta_W\sigma_{el}^{(\ell\bar{\ell})}
     \,,\quad T\gg T_c \,,     \label{sTT}
\end{eqnarray}
where the $Y_{q_{R,L}}$ are the right and left-handed quark hypercharges,
and the $Y_{R,L}$ are the right and left-handed charged lepton hypercharges.
The terms entering the prefactor are:  i) a factor 1/2 because only the
right-handed leptons contribute significantly to the conductivity; ii) a net
factor of $\cos^2\theta_W$ because the $B$ field coupling to right-handed
leptons and the current $J_B$ contain factors of $g'=e/\cos\theta_W$, while
the square of the matrix element contains a factor $(e/\cos\theta_W)^4$; and
iii) a factor $(Y_R^2+2Y_L^2)/8 = 3/4$ in the scatterings of the right-handed
charged leptons with right and left-handed leptons, and iv) a factor
$\sum_q^{N_q}(Y^2_{q_R}+Y^2_{q_L})/8 =11/12$ in the
scatterings of the right-handed charged leptons with right and left-handed
quarks.  
The factor 9/19 holds in the limit that the temperature is much greater than
the top quark mass, $m_t$; excluding the top quark for $T_c<T<m_t$ gives
108/211 instead.

    Applying a $W^3$ field to the electroweak plasma drives the charged
leptons and neutrinos oppositely since they couple through $g\tau_3W_3$.  In
this case, exchanges of $W^\pm$ dominate the interactions as charge is
transferred in the singular forward scatterings, so that ${\bf u}_3={\bf
u}_2=-{\bf u}_1$.  The collision term is then weighted by a factor $({\bf
p}_1+{\bf p}_2)$ instead of a factor $({\bf p}_1-{\bf p}_2)={\bf q}$ and one
ends up with an integral $\int p^2dq^2/(q^2+q_D^2)^2\simeq
T^2/q_D^2\sim\alpha^{-1}$ for the longitudinal part of the interaction.  For
the transverse part of the interaction one encounters a logarithmic
singularity; while Landau damping is not sufficient to screen the interaction,
a magnetic mass, $m_{mag}$, will provide an infrared cutoff.  Besides the
logarithms, the factor $\alpha^{-1}$ remains and we expect that
\begin{eqnarray}
   \sigma_{WW} \sim \alpha\, \sigma_{BB}.
\end{eqnarray}
This effect of $W^\pm$ exchange is analogous to the way gluon exchange in
QCD gives strong stopping and reduces the ``color conductivity" significantly
\cite{color}; similar effects are seen in spin diffusion in Fermi liquids
\cite{BPbook}.

    The electrical conductivity is found from $\sigma_{BB}$ and $\sigma_{WW}$
by rotating the $B$ and $W^3$ fields and currents by the Weinberg angle; using
Eq.  (\ref{sdef}) we obtain,
\begin{eqnarray}
   \left(\begin{array}{c} J_A \\ J_{Z^0} \end{array} \right)  &=&
   {\cal R}(\theta_W)\sigma {\cal R}(-\theta_W)
      \left(\begin{array}{c} A \\ Z^0 \end{array} \right) \nonumber \\
      &=&
   \left(\begin{array}{cc}\sigma_{BB}\cos^2\theta_W+\sigma_{WW}\sin^2\theta_W
     & \quad(\sigma_{BB}-\sigma_{WW})\cos\theta_W\sin\theta_W \\
       (\sigma_{WW}-\sigma_{BB})\cos\theta_W\sin\theta_W &
      \quad \sigma_{BB}\sin^2\theta_W +\sigma_{WW}\cos^2\theta_W \end{array}
\right)
   \left(\begin{array}{c} A \\ Z^0 \end{array} \right).   \label{sigmarot}
\end{eqnarray}
Thus the electrical conductivity is given by
\begin{eqnarray}
   \sigma_{AA} =
        \sigma_{BB}\cos^2\theta_W + \sigma_{WW}\sin^2\theta_W;
\end{eqnarray}
$\sigma_{el}/T$ above the electroweak transition differs from that below
mainly by a factor $\sim\cos^4\theta_W\simeq 0.6$.

    In the wide temperature range we are considering the coupling constants in
fact run as $\alpha_i(Q)=\alpha_i(\mu)+b_i\ln(Q/\mu)$ where the coefficients
$b_i$ are found by renormalization group calculations\cite{HM,Wilczek}.  In a
high temperature plasma typical momentum transfers $Q$ are of order $q_D\sim
eT$.  The exact values employed for $Q$ is not important as the couplings only
increase logarithmically with temperature.  In the temperature range 1 to
10$^6$ GeV, $\alpha^{-1}$ varies from 130 to 123 and $\sin^2\theta_W$ from
0.21 to 0.28.

\section{Summary and Outlook}

    We have calculated the electrical and electroweak conductivities in the
early universe over a wide range of temperatures.  Typically,
$\sigma_{el}\simeq T/\alpha^2\ln(1/\alpha)$, where the logarithmic dependence
on the coupling constant arises from Debye and dynamical screening of small
momentum-transfer interactions.  In the quark-gluon plasma, at $T\gg
T_{QGP}\sim 150$ MeV, the additional stopping on quarks reduces the electrical
conductivity from that in the hadronic phase.  In the electroweak
symmetry-restored phase, $T\gg T_c$, interactions between leptons and $W^\pm$
and $Z^0$ bosons reduce the conductivity further.  The electrical conductivity
does not vanish (as one might have imagined to result from singular unscreened
$W^\pm$-exchanges), and is larger than previous estimates, within an order of
magnitude.  The current is carried mainly by right-handed leptons since they
interact only through exchange of $\gamma$ and $Z^0$.

    From the above analysis we can infer the qualitative behavior of other
transport coefficients.  The characteristic electrical relaxation time,
$\tau_{el}\sim (\alpha^2\ln(1/\alpha)T)^{-1}$, defined from $\sigma\simeq e^2
n\tau_{el}/T$, is a typical ``transport time" which determines relaxation of
transport processes when charges are involved.  Right-handed leptons interact
through $Z^0$ exchanges only, whereas left-handed leptons may change into
neutrinos by $W^\pm$ exchanges as well.  Since $Z^0$ exchange is similar to
photon exchange when $T\gg T_c$, the characteristic relaxation time is similar
to that for electrical conduction, $\tau_{\nu}\sim
(\alpha^2\ln(1/\alpha)T)^{-1}$ (except for the dependence on the Weinberg
angle).  Thus the viscosity is $\eta\sim \tau_\nu \sim
T^3/(\alpha^2\ln(1/\alpha))$.  For $T\ll M_W$ the neutrino interaction is
suppressed by a factor $(T/M_W)^4$; in this regime neutrinos have longest mean
free paths and dominate the viscosity.\cite{BHP}

    The electrical conductivity of the plasma in the early universe is
sufficiently large that large-scale magnetic flux present in this period does
not diffuse significantly over timescales of the expansion of the universe.
The time for magnetic flux to diffuse on a distance scale $L$ is $\tau_{diff}
\sim \sigma_{el} L^2$.  Since the expansion timescale $t_{exp}$ is $\sim
1/(t_{\rm Planck}T^2)$, where $t_{\rm Planck} \sim 10^{-43}$ s is the Planck
time, one readily finds that
\begin{eqnarray}
\frac{\tau_{diff}}{t_{exp}} \sim
\alpha x^2 \frac{\tau_{el}}{t_{\rm Planck}} \gg 1,
\end{eqnarray}
where $x = L/ct_{exp}$ is the diffusion length scale in units of the
distance to the horizon.  As described in Refs.  \cite{Enqvist} and
\cite{LMcL}, sufficiently large domains with magnetic fields in the early
universe would survive to produce the primordial magnetic fields observed
today.

    We grateful to L.McLerran, C.J.  Pethick, J. Popp and B. Svetitsky for
discussion.  This work was supported in part by DOE Grant DE-AC03-76SF00098
and NSF Grant PHY 94-21309.

\vspace{2cm}

\figure{{\bf Fig. 1}:  The particle currents generated by an electric
field in the early universe at $T\gg T_c$.  Right-handed charged leptons,
$l=e,\mu\tau$,
interact only through $\gamma$ and $Z^0$ exchanges while left-handed leptons
also interact by $W^\pm$ exchanges, which drag neutrinos along and
decrease the current.  The vector and Higgs bosons, $\gamma, Z^0, W^\pm,\phi$,
cannot flow due to $W^\pm$ exchanges, and $q\bar{q}g$ are stopped by strong
interactions.
\label{Figscurrent}}

\figure{{\bf Fig. 2}:  Electrical conductivity vs. temperature.  The
temperatures, where the transitions from hadronic to quark-gluon plasma and
electroweak symmetry breaking occur, are indicated by QGP and EW respectively.
The conductivity $\sigma_{el}$ is given by Eqs.  (\ref{ec},\ref{eq},\ref{sTT})
in the three regions and are extrapolated into the regions of the phase
transitions.  The quark and lepton masses in the figure indicate the
temperatures at which they are thermally produced and thus affect the
conductivity (see Eq.  (\ref{eq}) and discussion in text).  \label{Figshydro}}

\newpage

\begin{figure}
\centerline{
\psfig{figure=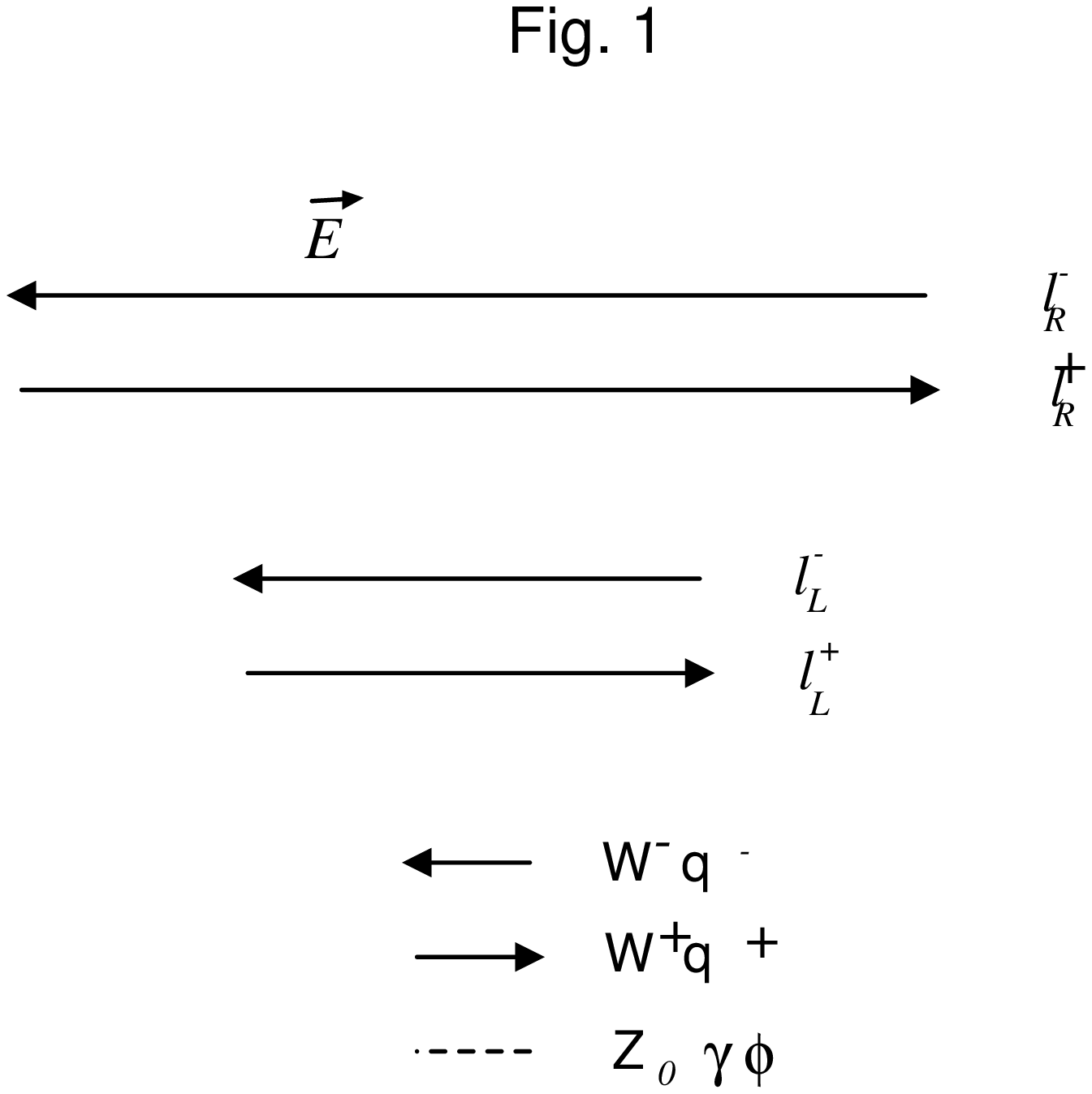,width=16cm,height=20cm,angle=-90}}
\vspace{1cm}
\caption{
}
\end{figure}

\newpage

\begin{figure}
\centerline{
\psfig{figure=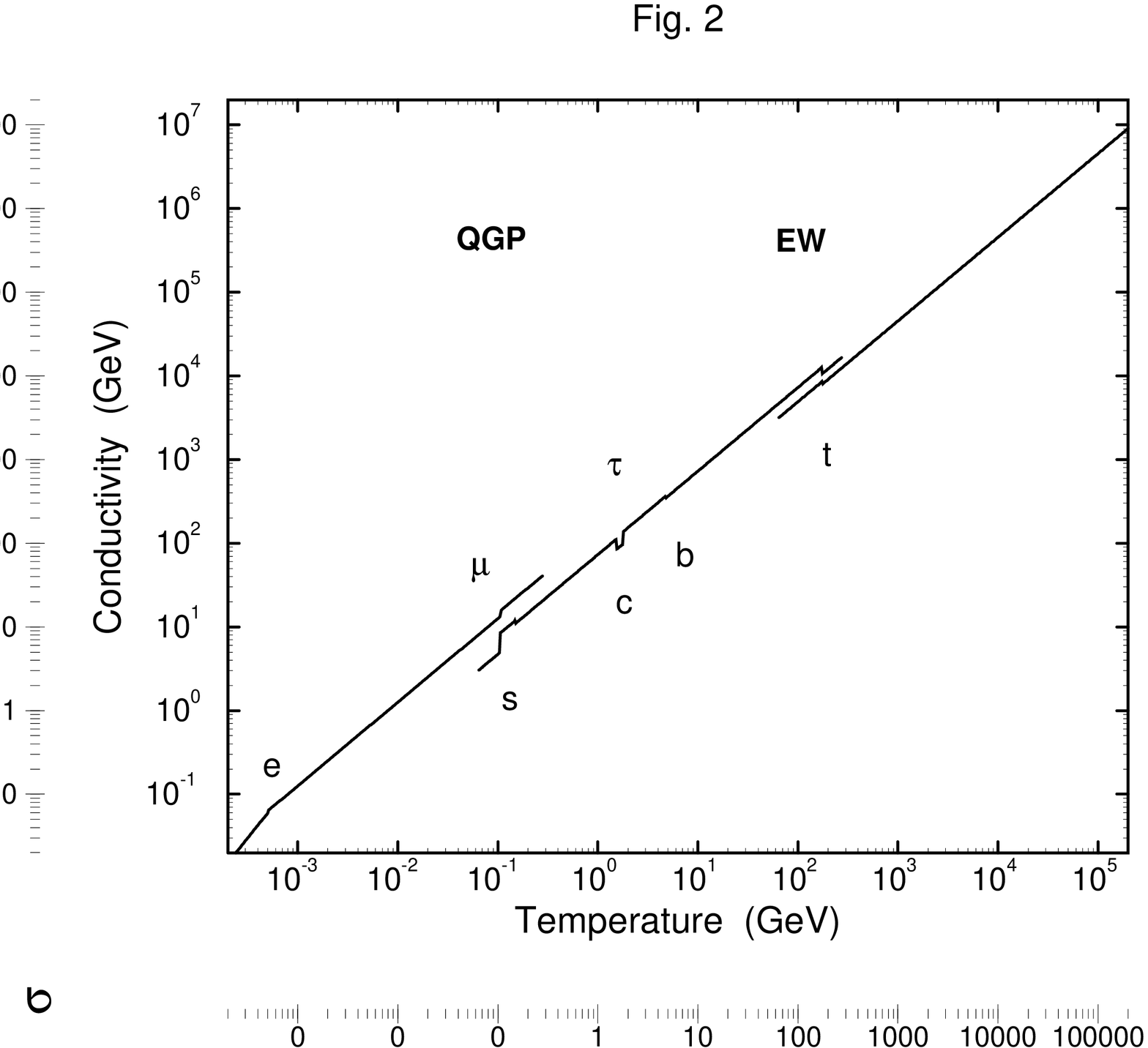,width=18cm,height=22cm,angle=-90}}
\vspace{1cm}
\caption{
}
\end{figure}

\end{document}